\documentclass[aps,pra,superscriptaddress,twocolumn]{revtex4-1}%
\usepackage{subfigure}
\usepackage{xcolor}
\usepackage{graphicx}
\usepackage{bm}
\usepackage{amsmath}
\usepackage{float}
\usepackage{ulem}
\usepackage[colorlinks,linkcolor=blue,anchorcolor=blue,citecolor=blue,urlcolor=blue]{hyperref}
\usepackage{multirow}

\newcommand{\dd}{\mathrm{d}}
\begin{document}
\title{Ground State and Collective Modes of Bose-Einstein Condensates in Newtonian and MOND-inspired gravitational potentials}
\author{Ning Liu}
\email{ningliu@mail.bnu.edu.cn}
\affiliation{School of Mathematics and Physics, Anqing Normal University, Anqing 246133, China}
\affiliation{Institute of Astronomy and Astrophysics, Anqing Normal University, Anqing 246133, China}
\affiliation{Key Laboratory of Multiscale Spin Physics(Ministry of Education),Beijing Normal University, Beijing 100875, China}


\begin{abstract}
We analytically and numerically study the ground state and collective dynamics of Bose-Einstein condensates in two traps: a Newtonian potential and a logarithmic potential inspired by Modified Newtonian Dynamics (MOND). In the ground state, the MOND potential supports bound states only in the deep-MOND regime, where the condensate becomes significantly larger than its Newtonian counterpart. The size increases with repulsive coupling parameter $\beta$ in both potentials. A clear scaling law of the size with $\beta^{1/3}$ emerges in the MOND case and is confirmed numerically over a wide parameter range, while for the Newtonian potential no simple scaling law exists as the Thomas-Fermi approximation ceases to be valid. For the dynamics, we derive and solve equations for the monopole collective mode. The larger MOND-bound condensate oscillates at a lower frequency, which scales as $\beta^{-1/3}$ in the strong-interaction limit. These scaling laws provide insights for quantum-simulation experiments aiming to probe modified-gravity scenarios with cold atoms.
\end{abstract}
\maketitle

\section{Introduction}
The nature of dark matter remains one of the most pressing puzzles in modern physics. It was originally proposed to explain galactic rotation curves and other gravitational anomalies on galactic and cosmological scales~\cite{arkani2009theory,bertone2018history}. Despite its success in addressing a wide range of astrophysical observations, dark matter has not been directly detected in laboratory experiments~\cite{bo2025dark,aalbers2025dark,zimmermann2025dwarf}. As an alternative paradigm, Modified Newtonian Dynamics (MOND) modifies Newtonian dynamics at low accelerations, offering an explanation for rotation curves without invoking dark matter~\cite{milgrom1983modification}. Notably, the MOND acceleration scale $a_0$ coincides with cosmologically relevant accelerations, suggesting a deep connection between MOND and cosmology~\cite{milgrom2014mond,bugg2015mond}. However, the microscopic foundation of MOND and its experimental verification at laboratory scales remain challenging~\cite{bekenstein1984does,sanders2002modified,bekenstein2004relativistic,famaey2012modified,bilek2018mond,milgrom2023generalizations,milgrom2025mond}.

Ultracold atomic gases, particularly Bose-Einstein condensates (BECs), have emerged as versatile platforms for quantum simulation due to their high controllability and precision measurement capabilities~\cite{dalfovo1999theory,morsch2006dynamics,pethick2008bose,bloch2008many,gross2017quantum}. These systems enable the emulation of various gravitational scenarios, including analog black holes~\cite{garay2000sonic,barcelo2011analogue,steinhauer2014observation,steinhauer2016observation,
munoz2019observation,barcelo2019analogue,kolobov2021observation}, dark matter~\cite{boehmer2007can,das2015dark,chavanis2017dissipative,harko2019jeans,chavanis2025review}, and cosmological models~\cite{eckel2018rapidly,viermann2022quantum,mondal2025dynamics}. The possibility of realizing gravity-like interactions in BEC experiments has been extensively explored~\cite{o2000bose,giovanazzi2001self,choi2002collision,chalony2013long,paredes2020optics,cartarius2008dynamics}, and systems with engineered long-range dipolar interactions have already been implemented~\cite{lu2011strongly,defenu2023long,bigagli2024observation}. This demonstrates the potential of BECs to probe fundamental gravitational physics in the laboratory. In this context, BECs offer a unique opportunity to test MOND-inspired gravitational potentials at microscopic scales, thanks to their ability to tailor external potentials and interaction strengths, combined with real-time imaging of density distributions and dynamics.

In this work, we systematically study the ground-state properties and collective excitations of a BEC trapped in two spherically symmetric gravitational potentials: the conventional Newtonian potential and a logarithmic potential derived from the deep-MOND regime. Using variational methods, we show that bound states exist for the MOND potential only in the deep-MOND regime, where the condensate becomes significantly larger and more weakly bound than in the Newtonian case. In the strong-interaction limit, we uncover a clean scaling law for the equilibrium size in the MOND potential, while the Newtonian counterpart exhibits no such simple scaling due to the breakdown of the Thomas-Fermi (TF) approximation. For the collective dynamics, we find that the monopole oscillation frequency in the MOND potential is notably lower in the deep-MOND regime and follows a distinct scaling law in the strong-coupling limit, providing a clear experimental signature to distinguish between the two gravitational models. Our results establish theoretical benchmarks for future quantum simulation experiments aimed at probing modified-gravity scenarios with ultracold atoms.

The paper is structured as follows: Section \ref{TM} outlines the theoretical model,which includes the Gross–Pitaevskii equation, the two distinct gravitational potentials (Newtonian and MOND-inspired), and the dimensionless scheme. Section \ref{G} presents analytical and numerical results for the ground-state energy, and demonstrates the scaling laws satisfied by the MOND potential within the TF approximation. Section \ref{DB} investigates the collective dynamics of the condensate; we derive the equations for the monopole breathing mode and numerically simulate its time evolution. Finally, Section \ref{co} summarizes the main findings and discusses their implications for future experiments.

\section{Theoretical Model}\label{TM}

\subsection{Gross-Pitaevskii Equation}

We consider a BEC in an external potential $V_{\text{ext}}(\mathbf{r})$. Within the mean-field approximation, the system's dynamics are described by the Gross-Pitaevskii (GP) equation. The energy functional can be written as:
\begin{equation}
E[\psi] = \int d^3\mathbf{r} \bigg[ \frac{\hbar^2}{2m} |\nabla\psi(\mathbf{r})|^2
+ V_{\text{ext}}(\mathbf{r}) |\psi(\mathbf{r})|^2+ \frac{g}{2} |\psi(\mathbf{r})|^4 \bigg],
\label{eq:energy_functional}
\end{equation}
where $m$ is the atomic mass, $g = 4\pi\hbar^2 a_s/m$ is the contact interaction strength. Here, $a_s$ is the $s$-wave scattering length. The wavefunction satisfies the normalization condition $\int |\psi|^2 d^3\mathbf{r} = N$, where $N$ is the total number of atoms. From the energy functional, the variational principle $i\hbar\partial_t \psi = \delta E/\delta \psi^*$ yields the time-dependent GP equation~\cite{pitaevskii2016bose}:
\begin{equation}
i\hbar\frac{\partial \psi(\mathbf{r},t)}{\partial t} = \left[-\frac{\hbar^2}{2m}\nabla^2 + V_{\text{ext}}(\mathbf{r}) + g|\psi(\mathbf{r},t)|^2\right] \psi(\mathbf{r},t).
\label{eq:time_gp}
\end{equation}

For the stationary case, the wavefunction can be written as $\psi(\mathbf{r},t) = \phi(\mathbf{r})e^{-i\mu t/\hbar}$. Substituting into Eq.(\ref{eq:time_gp}) gives the stationary GP equation:
\begin{equation}
\left[-\frac{\hbar^2}{2m}\nabla^2 + V_{\text{ext}}(\mathbf{r}) + g|\phi(\mathbf{r})|^2\right] \phi(\mathbf{r}) = \mu \phi(\mathbf{r}),
\label{eq:stationary_gp}
\end{equation}
where $\mu$ is the chemical potential.

This paper considers two spherically symmetric gravitational potentials for $V_{\text{ext}}(\mathbf{r})$. In classical Newtonian gravity, the gravitational potential generated by a mass $M$ is:
\begin{equation}
\Phi_N(r) = -\frac{GM}{r},
\label{eq:newton_potential_phi}
\end{equation}
where $G$ is the gravitational constant. The potential energy felt by an atom in the BEC is:
\begin{equation}
V_N(r) = m\Phi_N(r) = -\frac{GMm}{r}.
\label{eq:newton_potential}
\end{equation}
In MOND theory, the gravitational potential in the deep-MOND regime ($a \ll a_0$) is governed by the modified Poisson Eq.~\cite{bekenstein1984does}:
\begin{equation}
\nabla \cdot \left[ \mu\!\left( \frac{|\nabla \Phi|}{a_0} \right) \nabla \Phi \right] = 4\pi G \rho,
\label{eq:mond_poisson}
\end{equation}
where $a_0 \approx 1.2 \times 10^{-10} \text{m/s}^2$ is the MOND characteristic acceleration, $\Phi$ is the gravitational potential, $\rho$ is the mass density of the atomic cloud, and $\mu(x)$ is the MOND interpolation function, which satisfies:
\begin{equation}
\mu(x) = 
\begin{cases}
x, & x \ll 1 \quad (\text{deep MOND region}) \\
1. & x \gg 1 \quad (\text{Newtonian region})
\end{cases}
\end{equation}
Solving Eq.(\ref{eq:mond_poisson}) in the deep MOND region yields the gravitational potential~\cite{bugg2015mond}:
\begin{equation}
\Phi_M(r) = \sqrt{GMa_0} \ln\left(\frac{r}{r_M}\right)
\label{eq:mond_potential_phi}
\end{equation}
where $r_M = \sqrt{GM/a_0}$ is the characteristic distance, satisfying $\Phi_M(r_M) = 0$. Thus, the potential energy felt by an atom in the BEC is:
\begin{equation}
V_M(r) = m\Phi_M(r) = m\sqrt{GMa_0} \ln\left(\frac{r}{r_M}\right)
\label{eq:mond_potential}
\end{equation}

\subsection{Dimensionless Scheme}\label{non-dimensionalization}

To highlight the essential physics and simplify the calculations, we introduce the following dimensionless scheme. We define the characteristic length $l_0$ as the scale at which the quantum pressure balances Newtonian gravity:
\begin{equation}
l_0 = \frac{\hbar^2}{GMm^2}.
\end{equation}
The corresponding characteristic energy is:
\begin{equation}
E_0 = \frac{GMm}{l_0} = \frac{(GM)^2 m^3}{\hbar^2}.
\end{equation}
We then define the dimensionless variables:
\begin{align}
\tilde{r} &= \frac{r}{l_0}, \quad \tilde{\sigma} = \frac{\sigma}{l_0}.
\end{align}
The dimensionless forms of the potential energies, $\tilde{V}_N(\tilde{r}) = V_N/E_0$ and $\tilde{V}_M(\tilde{r}) = V_M/E_0$, are thus
\begin{align}
\tilde{V}_N(\tilde{r}) &= -\frac{1}{\tilde{r}}, \\
\tilde{V}_M(\tilde{r}) &= \eta \ln\left(\tilde{r}\eta\right),
\label{eq:corrected_mond_potential}
\end{align}
where $\eta = l_0/r_M$ is the MOND parameter that characterizes the strength of the MOND effect,
\begin{equation}
\eta = \frac{\hbar^2 \sqrt{a_0}}{(GM)^{3/2} m^2}.
\end{equation}
Physically, $\eta \ll 1$ corresponds to the deep-MOND regime, $\eta \gg 1$ to the Newtonian regime, and $\eta \sim 1$ to the crossover region. In what follows, we systematically compare the effects of the MOND and Newtonian potentials on condensate properties from both ground-state and dynamical perspectives. We also introduce the dimensionless coupling strength $\beta = gN/(E_0 l_0^3)$, which can be expressed as:
\begin{equation}
\beta = \frac{4\pi a_s GMm^2 N}{\hbar^4}.
\end{equation}
In cold-atom experiments, the s-wave scattering length $a_s$ can be tuned over a wide range via Feshbach resonance, allowing flexible control of the interaction parameter $\beta$.

\section{Ground State Properties}\label{G}
\subsection{Ground Energy}\label{GE}
We investigate the ground state of BECs under the gravitational potentials given by Eqs.~(\ref{eq:newton_potential}) and (\ref{eq:mond_potential}). Owing to the spherical symmetry of the potentials, we employ a Gaussian trial wavefunction,
\begin{equation}
\psi_0(r) = \sqrt{\frac{N}{(\pi \sigma^2)^{3/2}}} \exp\left(-\frac{r^2}{2\sigma^2}\right),
\label{eq:gaussian_ground_state}
\end{equation}
where $\sigma$ is the variational parameter representing the width of the wavefunction. Substituting this ansatz into the energy functional Eq.~(\ref{eq:energy_functional}) yields the total energy
\begin{equation}
E(\sigma) = T + V + E_{\text{int}},
\label{eq:total_energy_sigma}
\end{equation}
with the kinetic energy term
\begin{equation}
T = \frac{3\hbar^2 N}{4m\sigma^2},
\label{eq:kinetic_energy_ground}
\end{equation}
the potential energy term
\begin{equation}
V = N \int_0^\infty V_{\text{ext}}(r) \frac{4\pi r^2}{(\pi\sigma^2)^{3/2}} e^{-r^2/\sigma^2} dr,
\label{eq:potential_energy_ground}
\end{equation}
and the interaction energy
\begin{equation}
E_{\text{int}} = \frac{gN^2}{2(2\pi)^{3/2} \sigma^3}.
\label{eq:interaction_energy_ground}
\end{equation}
For the Newtonian potential $V_N(r)$, the integral in Eq.~(\ref{eq:potential_energy_ground}) evaluates to
\begin{equation}
V_N = -\frac{GMmN}{\sigma} \sqrt{\frac{4}{\pi}}.
\label{eq:newton_potential_energy}
\end{equation}
For the MOND potential $V_M(r)$, we obtain
\begin{equation}
V_M = m\sqrt{GMa_0} N \left[ \ln\left(\frac{\sigma}{r_M}\right) + 1 - \frac{\gamma}{2} - \ln 2 \right],
\label{eq:mond_potential_energy}
\end{equation}
where $\gamma \approx 0.5772$ is the Euler constant.

Using the dimensionless scheme introduced in Section~\ref{non-dimensionalization}, the dimensionless energy per particle for the Newtonian potential reads
\begin{equation}
\tilde{e}_N(\tilde{\sigma}) = \frac{3}{4\tilde{\sigma}^2} - \sqrt{\frac{4}{\pi}}\frac{1}{\tilde{\sigma}} + \frac{\beta}{2(2\pi)^{3/2}}\frac{1}{\tilde{\sigma}^3}.
\label{eq:dimless_per_particle_energy_newton}
\end{equation}
For the MOND potential, the corresponding expression is
\begin{equation}
\begin{aligned}
\tilde{e}_M(\tilde{\sigma}) &= \frac{3}{4\tilde{\sigma}^2} + \eta\left(\ln\tilde{\sigma} + \ln\eta + 1 - \frac{\gamma}{2} - \ln 2\right)\\ &+ \frac{\beta}{2(2\pi)^{3/2}}\frac{1}{\tilde{\sigma}^3}.
\end{aligned}
\label{eq:dimless_per_particle_energy_mond_corrected}
\end{equation}

Minimizing the energy with respect to $\tilde{\sigma}$ provides the ground state. For the Newtonian potential, setting $\dd\tilde{e}_N/\dd\tilde{\sigma} = 0$ gives
\begin{equation}
-\frac{3}{2}\tilde{\sigma} + \sqrt{\frac{4}{\pi}}\tilde{\sigma}^2 - \frac{3\beta}{2(2\pi)^{3/2}} = 0.
\label{eq:derivative_newton}
\end{equation}
For the MOND potential, $\dd\tilde{e}_M/\dd\tilde{\sigma}=0$ yields
\begin{equation}
-\frac{3}{2}\tilde{\sigma} + \eta\tilde{\sigma}^3 - \frac{3\beta}{2(2\pi)^{3/2}} = 0.
\label{eq:derivative_mond}
\end{equation}
Solving these equations numerically yields the equilibrium width $\tilde{\sigma}_0$ and the corresponding ground-state energy $\tilde{e}_0$.

\begin{figure}[!ht]
\centering
\includegraphics[width=0.45\textwidth]{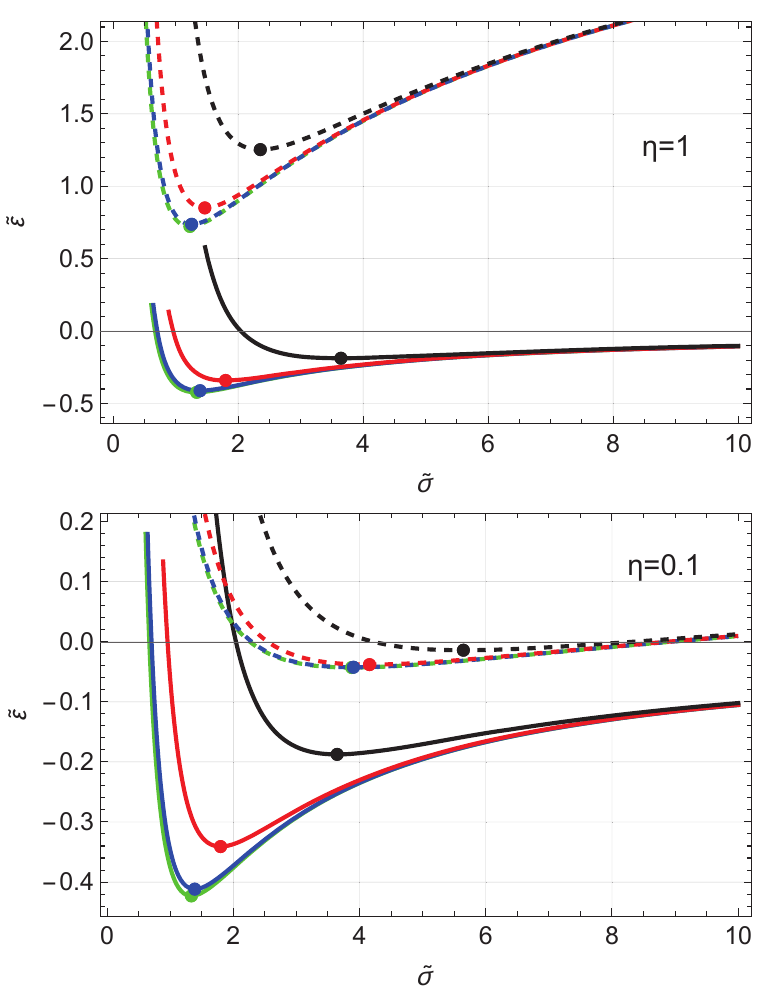}
\caption{Dimensionless energy per atom $\tilde{e}$ as a function of the normalized width $\tilde{\sigma}$ for the Newtonian (solid lines) and MOND (dashed lines) potentials. Different colors correspond to different interaction strengths: $\beta=0.1$ (green), $1$ (blue), $10$ (red), and $100$ (black). Energy minima are marked by circles of the corresponding color.}
\label{fig:ground_state_energy}
\end{figure}

Figure~\ref{fig:ground_state_energy} shows the dimensionless energy per atom $\tilde{e}$ as a function of the normalized width $\tilde{\sigma}$ for different values of $\beta$ and $\eta$. A key observation is the distinct behavior between the two regimes of the MOND parameter. For $\eta=1$ (crossover regime), the energy curves for the MOND potential lie above zero, indicating the absence of a bound state. In contrast, for $\eta=0.1$ (deep-MOND regime), the MOND potential supports a negative energy minimum, allowing bound-state formation. However, the depth of this minimum is significantly shallower than that of the Newtonian potential at the same $\beta$, indicating a much weaker confinement. 

Furthermore, in the deep-MOND regime, the equilibrium width increases with $\beta$, and the condensate is notably broader than its Newtonian counterpart at the same interaction strength. As the deep-MOND regime is precisely where MOND effects become dominant, this leads to a key conclusion: compared to the Newtonian potential, the MOND potential tends to produce a more spatially extended and more weakly bound condensate.

The numerical values of the equilibrium width $\tilde{\sigma}_0$ and ground-state energy $\tilde{e}_0$, summarized in Table~\ref{tab:ground_state_properties}, confirm and quantify the qualitative picture from Fig.~\ref{fig:ground_state_energy}. Specifically, for $\eta=1$ the MOND ground-state energy $\tilde{e}_0^M$ is positive for all $\beta$, precluding bound states, while for $\eta=0.1$ it becomes negative—confirming bound-state formation—yet with binding energies roughly an order of magnitude smaller than the Newtonian counterparts $\tilde{e}_0^N$, consistent with a much weaker confinement. The table further shows that in the deep-MOND regime ($\eta=0.1$) the condensate width $\tilde{\sigma}_0^M$ is significantly larger than $\tilde{\sigma}_0^N$ at each $\beta$, directly supporting the conclusion of a more extended, weakly bound condensate under MOND.
\begin{table}[!ht]
\centering
\caption{Equilibrium width $\tilde{\sigma}_0$ and ground-state energy $\tilde{e}_0$ for the Gaussian trial wavefunction, for different values of the MOND parameter $\eta$ and the interaction strength $\beta$. Superscripts $N$ and $M$ denote Newtonian and MOND potentials, respectively.}
\begin{tabular}{ccccccc}
\hline
$\eta$ & $\beta$ & $\tilde{\sigma}_0^N$ & $\tilde{e}_0^N$ & $\tilde{\sigma}_0^M$ & $\tilde{e}_0^M$ & $\tilde{\sigma}_0^M/\tilde{\sigma}_0^N$ \\
\hline
\multirow{4}{*}{1}
& 0.1  & 1.33566 & -0.423071 & 1.22791 & 0.722699 & 0.919326 \\
& 1    & 1.39006 & -0.411784 & 1.25534 & 0.737624 & 0.90308  \\
& 10   & 1.79862 & -0.340961 & 1.46615 & 0.850519 & 0.815155 \\
& 100  & 3.64498 & -0.187563 & 2.35471 & 1.25309  & 0.646015 \\
\hline
\multirow{4}{*}{0.1}
& 0.1  & 1.33566 & -0.423071 & 3.87615 & -0.0429769 & 2.90205 \\
& 1    & 1.39006 & -0.411784 & 3.90435 & -0.0424916 & 2.80876 \\
& 10   & 1.79862 & -0.340961 & 4.15818 & -0.0381341 & 2.31188 \\
& 100  & 3.64498 & -0.187563 & 5.64539 & -0.0141724 & 1.54881 \\
\hline
\end{tabular}
\label{tab:ground_state_properties}
\end{table}

Beyond this verification, several important trends emerge from the quantitative data. First, the equilibrium width increases with $\beta$ for both potentials, reflecting the expansion driven by repulsive interactions. Notably, the relative size of the MOND condensate depends crucially on $\eta$: for $\eta=1$, it is more compact than its Newtonian counterpart ($\tilde{\sigma}_0^M < \tilde{\sigma}_0^N$), with the ratio $\tilde{\sigma}_0^M/\tilde{\sigma}_0^N$ decreasing from about 0.92 to 0.65 as $\beta$ increases from 0.1 to 100. In contrast, for $\eta=0.1$ (deep-MOND regime), the MOND condensate becomes substantially larger ($\tilde{\sigma}_0^M > \tilde{\sigma}_0^N$), with the width ratio decreasing from about 2.90 to 1.55 over the same $\beta$ range. This reversal underscores the pivotal role of the MOND parameter $\eta$ in controlling the condensate's spatial extent.

The underlying mechanism can be traced to the form of the gravitational potential. In the deep-MOND regime, the potential is logarithmic, $V_M(r) \propto \ln r$, resulting in a weaker restoring force (proportional to $1/r$) compared to the Newtonian case ($\propto 1/r^2$). This weaker confinement necessitates a larger spatial distribution to balance the quantum pressure and repulsive interactions, leading to a larger condensate size and a shallower bound state. This finding provides clear theoretical guidance for experiments aiming to observe the transition between bound and unbound states by tuning the effective MOND parameter $\eta$, for instance, through the control of the equivalent source mass $M$ in a laboratory simulation.

\subsection{Analytical Solutions and Scaling Law Analysis in the Strong Interaction Limit}\label{GA}

Within the Thomas-Fermi (TF) approximation, which neglects the kinetic energy in the strong interaction limit ($\beta \gg 1$), the energy functionals simplify considerably. For the Newtonian potential, the dimensionless energy per particle reduces to
\begin{equation}
\tilde{e}_N^{\mathrm{TF}}(\tilde{\sigma}) = -\sqrt{\frac{4}{\pi}}\frac{1}{\tilde{\sigma}} + \frac{\beta}{2(2\pi)^{3/2}}\frac{1}{\tilde{\sigma}^3}.
\label{eq:strong_energy_newton}
\end{equation}
Minimizing this expression with respect to $\tilde{\sigma}$ ($\dd\tilde{e}_N^{\mathrm{TF}}/\dd\tilde{\sigma}=0$) yields the equilibrium width
\begin{equation}
\tilde{\sigma}_0^{N,\mathrm{TF}} = \left( \frac{3\beta}{2\sqrt{2\pi}} \right)^{1/2},
\label{eq:approx_solution_newton}
\end{equation}
implying a scaling law $\tilde{\sigma}_0^N \propto \beta^{1/2}$. However, numerical results reveal that this approximation fails even for moderately large $\beta$, indicating that the TF approximation is not applicable for the Newtonian potential in this regime. For instance, at $\beta=100$, Eq.~(\ref{eq:approx_solution_newton}) gives $\tilde{\sigma}_0^{N,\mathrm{TF}} \approx 19.3$, which is an order of magnitude larger than the exact numerical result $\tilde{\sigma}_0^N = 3.64$ (see Table~\ref{tab:ground_state_properties}). This failure stems from the competing roles of the different energy terms. In the exact energy functional, the kinetic term scales as $1/\tilde{\sigma}^2$, the Newtonian gravitational term as $-1/\tilde{\sigma}$, and the interaction term as $\beta/\tilde{\sigma}^3$. In the equilibrium condition derived from the full functional, the kinetic and gravitational terms remain comparable across a wide range of $\beta$, and the interaction term acts only as a perturbation that moderately increases the width. The TF approximation, by completely neglecting the kinetic term, overestimates the influence of the interaction and thus predicts an unrealistically large width that diverges from the true solution.

\begin{figure}[!ht]
\centering
\includegraphics[width=0.48\textwidth]{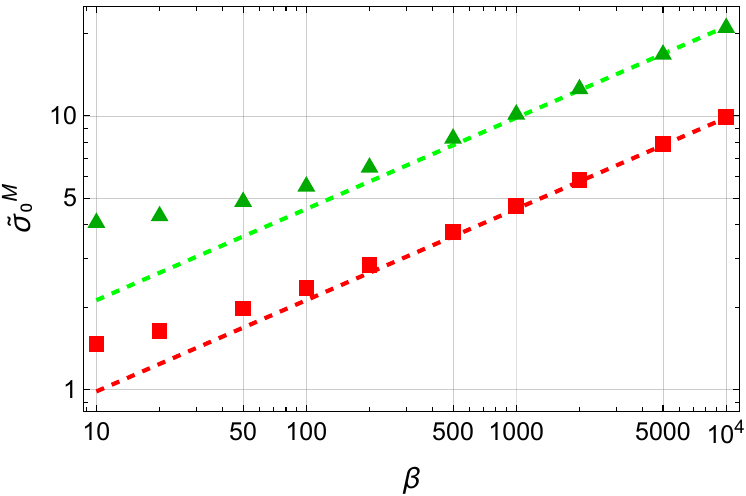}
\caption{Scaling of the condensate width $\tilde{\sigma}_0^M$ with the interaction parameter $\beta$ in the MOND potential. The dashed lines show the theoretical scaling $\tilde{\sigma}_0^M \propto \beta^{1/3}$ for $\eta=1$ (red) and $\eta=0.1$ (green). The symbols represent exact numerical solutions, demonstrating excellent agreement with the theoretical prediction across the entire range $\beta=10^1 \sim 10^4$.}
\label{fig:mond_scaling_laws}
\end{figure}

In sharp contrast to the Newtonian case, the TF approximation proves highly accurate for the MOND potential. Neglecting the kinetic energy, the energy functional simplifies to
\begin{equation}
\begin{aligned}
\tilde{e}_M^{\mathrm{TF}}(\tilde{\sigma}) &= \eta\left(\ln\tilde{\sigma} + \ln\eta + 1 - \frac{\gamma}{2} - \ln 2\right) \\ &+ \frac{\beta}{2(2\pi)^{3/2}}\frac{1}{\tilde{\sigma}^3}.
\end{aligned}
\label{eq:strong_energy_mond}
\end{equation}
Minimization ($\dd\tilde{e}_M^{\mathrm{TF}}/\dd\tilde{\sigma}=0$) yields the analytical solution for the equilibrium width
\begin{equation}
\tilde{\sigma}_0^{M,\mathrm{TF}} = \left( \frac{3\beta}{2(2\pi)^{3/2}\eta} \right)^{1/3},
\label{eq:exact_solution_mond}
\end{equation}
and the corresponding ground-state energy
\begin{equation}
\begin{aligned}
\tilde{e}_0^{M,\mathrm{TF}} &= \eta\left[\frac{1}{3}\ln\beta + \frac{2}{3}\ln\eta + \frac{1}{3}\ln\left( \frac{3}{2(2\pi)^{3/2}} \right) \right.\\&\left.+ \frac{4}{3} - \frac{\gamma}{2} - \ln 2\right].
\label{eq:exact_energy_mond}
\end{aligned}
\end{equation}
Eq.(\ref{eq:exact_solution_mond}) immediately implies the scaling law $\tilde{\sigma}_0^M \propto \beta^{1/3}$. Crucially, this analytical result agrees excellently with full numerical solutions across a wide range of $\beta$, as clearly shown in Fig.~\ref{fig:mond_scaling_laws}. The physical reason for this agreement lies in the scaling of the energy derivatives. For the MOND potential, the derivative of the kinetic term scales as $1/\tilde{\sigma}^3$, while that of the gravitational term scales as $\eta/\tilde{\sigma}$. Their ratio, $\propto 1/(\eta\tilde{\sigma}^2)$, decreases rapidly as $\tilde{\sigma}$ increases with $\beta$. Consequently, the kinetic energy contribution becomes negligible in the strong interaction limit, validating the TF approximation.

The distinct behaviors in the strong interaction limit stem from the fundamentally different forms of the gravitational potentials and their interplay with the kinetic energy. For the logarithmic MOND potential $V_M(r) \propto \ln r$, the TF approximation yields an equilibrium condition that balances the gravitational term derivative $\eta/\tilde{\sigma}$ against the interaction term derivative $\beta/\tilde{\sigma}^4$. This balance directly leads to the clean scaling relation $\tilde{\sigma}^3 \propto \beta/\eta$, i.e., $\tilde{\sigma} \propto \beta^{1/3}$. In stark contrast, for the Newtonian potential $V_N(r) \propto -1/r$, a similar analysis that naively balances the gravitational term derivative ($\propto 1/\tilde{\sigma}^2$) with the interaction term derivative ($\propto \beta/\tilde{\sigma}^4$) would suggest $\tilde{\sigma}^2 \propto \beta$. However, as our earlier analysis demonstrated, this simplistic balance fails because the kinetic term (with derivative $\propto 1/\tilde{\sigma}^3$) remains comparable to the gravitational term and cannot be neglected. Consequently, the Newtonian case does not exhibit a simple, pure power-law scaling; instead, the condensate width increases with $\beta$ in a more complex manner that depends on the interplay of all three energy terms.

These contrasting behaviors provide a decisive experimental signature to discriminate between the gravitational models. By exploiting Feshbach resonance to tune the s-wave scattering length $a_s$, and thus the parameter $\beta$, one can measure the condensate size as a function of $\beta$. Plotting the data on a double-logarithmic scale and extracting the slope offers a direct test: a robust and constant slope near $1/3$ supports the MOND-type logarithmic potential. In contrast, a slope that is not constant and varies significantly with $\beta$ would be indicative of the Newtonian $1/r$ potential, where kinetic energy effects preclude a simple scaling law. For a typical $^{87}\mathrm{Rb}$ BEC experiment with $N\sim10^5$ and $l_0\sim1\ \mu\mathrm{m}$, $\beta$ can be tuned across the range $10^2$-$10^4$ via Feshbach resonance, amply covering the parameter space needed to test this prediction.

\section{Dynamical Behavior and Collective Excitation Spectrum}\label{DB}

\subsection{Time-Dependent Variational Method and Derivation of Equations of Motion}

To investigate the collective dynamics of the condensate, we employ the time-dependent variational method. For the spherically symmetric potentials considered here, we focus on the breathing (monopole) mode. A suitable trial wavefunction is the spherically symmetric Gaussian wavefunction~\cite{perez1996low,haas2018time}:
\begin{equation}
\psi(\mathbf{r},t) = \sqrt{\frac{N}{[\pi w(t)^2]^{3/2}}} \exp\left[-\frac{r^2}{2w(t)^2} + i\alpha(t) r^2\right],
\label{eq:spherical_gaussian_time}
\end{equation}
where $w(t)$ is the time-dependent width and $\alpha(t)$ is a time-dependent quadratic phase parameter. The Lagrangian density for the GP equation is
\begin{equation}
\mathcal{L} = \frac{i\hbar}{2}\left(\psi^*\partial_t\psi - \psi\partial_t\psi^*\right) - \frac{\hbar^2}{2m}|\nabla\psi|^2 - V_{\text{ext}}(r)|\psi|^2 - \frac{g}{2}|\psi|^4.
\end{equation}
Substituting the ansatz Eq.~(\ref{eq:spherical_gaussian_time}) and integrating over space yields the effective Lagrangian
\begin{align}
L = \int \mathcal{L}  d^3\mathbf{r}&= -\frac{3}{2}\hbar N \dot{\alpha} w^2 - \frac{3\hbar^2 N}{4m w^2} - \frac{3\hbar^2 N}{m} \alpha^2 w^2 \nonumber\\ 
  &- \langle V_{\text{ext}} \rangle - \frac{g N^2}{2(2\pi)^{3/2} w^3},
\label{eq:effective_lagrangian_full}
\end{align}
where $\langle V_{\text{ext}} \rangle = \int V_{\text{ext}}(r) |\psi|^2  d^3\mathbf{r}$ is the total external potential energy.

Applying the Euler-Lagrange equations to the variational parameters $\alpha$ and $w$ gives their equations of motion. For the phase parameter $\alpha$, this yields the relation
\begin{equation}
\alpha = \frac{m}{2\hbar} \frac{\dot{w}}{w}.
\label{eq:alpha_relation}
\end{equation}
For the width $w$, after substituting Eq.~(\ref{eq:alpha_relation}) and its time derivative, we obtain a closed equation:
\begin{equation}
\ddot{w} = \frac{\hbar^2}{m^2 w^3} - \frac{2}{3m} \frac{dU}{dw} + \frac{g N}{m (2\pi)^{3/2} w^4},
\label{eq:width_evolution_general_U}
\end{equation}
where $U(w) = \langle V_{\text{ext}} \rangle / N$ is the average external potential energy per atom.

\subsection{Specific Equations of Motion for the Two Gravitational Potentials}

The form of $U(w)$ and its derivative depend on the specific gravitational potential. For the Newtonian potential $V_N(r) = -GMm/r$, we find
\begin{equation}
U_N(w) = -\frac{GMm}{w}\sqrt{\frac{4}{\pi}}, \quad \frac{dU_N}{dw} = \frac{GMm}{w^2}\sqrt{\frac{4}{\pi}}.
\end{equation}
Substituting into Eq.~(\ref{eq:width_evolution_general_U}) gives the width evolution equation for the Newtonian case:
\begin{equation}
\ddot{w} = \frac{\hbar^2}{m^2 w^3} - \frac{2GM}{3w^2}\sqrt{\frac{4}{\pi}} + \frac{g N}{m (2\pi)^{3/2} w^4}.
\label{eq:newton_width_evolution}
\end{equation}
For the MOND potential $V_M(r) = m\sqrt{GMa_0} \ln(r/r_M)$, we have
\begin{equation}
\begin{aligned}
U_M(w) &= m\sqrt{GMa_0} \left[ \ln\left(\frac{w}{r_M}\right) + 1 - \frac{\gamma}{2} - \ln 2 \right], \\
\frac{dU_M}{dw} &= \frac{m\sqrt{GMa_0}}{w},
\end{aligned}
\end{equation}
leading to the MOND width evolution equation:
\begin{equation}
\ddot{w} = \frac{\hbar^2}{m^2 w^3} - \frac{2\sqrt{GMa_0}}{3w} + \frac{g N}{m (2\pi)^{3/2} w^4}.
\label{eq:mond_width_evolution}
\end{equation}

We adopt the same dimensionless scheme as in Sec.~\ref{non-dimensionalization}. Using the characteristic length $l_0 = \hbar^2/(GMm^2)$ and time $t_0 = \sqrt{l_0^3/(GM)}$, we define the dimensionless widths $\tilde{w}_N = w_N/l_0$, $\tilde{w}_M = w_M/l_0$, and dimensionless time $\tilde{t}=t/t_0$. We recall the dimensionless parameters: the interaction strength $\beta = gN/(E_0 l_0^3)$ and the MOND parameter $\eta = l_0/r_M$, with $E_0 = GMm/l_0$.

The dimensionless evolution equations are then:
\begin{align}
 & \frac{d^2\tilde{w}_N}{d\tilde{t}^2} = \frac{1}{\tilde{w}_N^3} - \frac{2}{3}\sqrt{\frac{4}{\pi}} \frac{1}{\tilde{w}_N^2} + \frac{\beta}{(2\pi)^{3/2} \tilde{w}_N^4}, \label{eq:dimless_newton_width_eq} \\
 & \frac{d^2\tilde{w}_M}{d\tilde{t}^2} = \frac{1}{\tilde{w}_M^3} - \frac{2}{3}\frac{\eta}{\tilde{w}_M} + \frac{\beta}{(2\pi)^{3/2} \tilde{w}_M^4}. \label{eq:dimless_mond_width_eq}
\end{align}
The key difference lies in the gravitational terms: $-\frac{2}{3}\sqrt{4/\pi}\,\tilde{w}_N^{-2}$ for Newtonian and $-\frac{2}{3}\eta\,\tilde{w}_M^{-1}$ for MOND, reflecting the $1/r^2$ versus $1/r$ force laws.

\subsection{Equilibrium Width and Collective Excitation Frequencies}

The equilibrium widths $\tilde{w}_{0N}$ and $\tilde{w}_{0M}$ satisfy $\ddot{\tilde{w}}=0$, which for the two potentials gives
\begin{align}
 & \frac{1}{\tilde{w}_{0N}^3} - \frac{2}{3}\sqrt{\frac{4}{\pi}}\frac{1}{\tilde{w}_{0N}^2} + \frac{\beta}{(2\pi)^{3/2}\tilde{w}_{0N}^4} = 0, \label{eq:newton_equilibrium_eq} \\
 & \frac{1}{\tilde{w}_{0M}^3} - \frac{2}{3}\frac{\eta}{\tilde{w}_{0M}} + \frac{\beta}{(2\pi)^{3/2}\tilde{w}_{0M}^4} = 0. \label{eq:mond_equilibrium_eq}
\end{align}
These equations are identical to the energy minimization conditions derived in Sec.~\ref{GE} (Eqs.~(\ref{eq:derivative_newton}) and (\ref{eq:derivative_mond})), confirming the consistency between the static and dynamical descriptions. Their solutions, $\tilde{w}_{0N}$ and $\tilde{w}_{0M}$, are functions of $\beta$ and $\eta$, and have been tabulated in Table~\ref{tab:ground_state_properties}.

The frequency of small-amplitude breathing oscillations around equilibrium is found by linearizing Eqs.~(\ref{eq:dimless_newton_width_eq}) and (\ref{eq:dimless_mond_width_eq}). For a perturbation $\delta\tilde{w} \propto e^{-i\tilde{\Omega}\tilde{t}}$, we obtain
\begin{align}
\tilde{\Omega}_N^2 &= \frac{2}{\tilde{w}_{0N}^4} - \frac{2}{3}\sqrt{\frac{4}{\pi}}\frac{1}{\tilde{w}_{0N}^3} + \frac{3\beta}{(2\pi)^{3/2}\tilde{w}_{0N}^5}, \label{eq:newton_frequency_final} \\
\tilde{\Omega}_M^2 &= \frac{2}{\tilde{w}_{0M}^4} - \frac{1}{3}\frac{\eta}{\tilde{w}_{0M}^2} + \frac{3\beta}{(2\pi)^{3/2}\tilde{w}_{0M}^5}. \label{eq:mond_frequency_final}
\end{align}
These expressions explicitly show how the collective excitation frequencies depend on the equilibrium widths and the system parameters. For the MOND potential, the term $-\eta/(3\tilde{w}_{0M}^2)$ directly links the oscillation frequency to the MOND parameter $\eta$. A smaller $\eta$ (deeper MOND regime) reduces this term, contributing to a lower frequency $\tilde{\Omega}_M$, consistent with a weaker logarithmic confinement.

In the strong interaction limit ($\beta\gg1$), the TF approximation becomes accurate for the MOND potential, as established in Sec.~\ref{GA}. In this limit, the equilibrium width simplifies to $\tilde{w}_{0M}^{\mathrm{TF}} = \left( 3\beta/[2(2\pi)^{3/2}\eta] \right)^{1/3}$. Substituting this into Eq.~(\ref{eq:mond_frequency_final}) and neglecting terms of order $\tilde{w}_{0M}^{-4}$, we obtain an explicit expression for the squared frequency:
\begin{equation}
\left(\tilde{\Omega}_M^{\mathrm{TF}}\right)^2 \propto \eta^{5/3} \beta^{-2/3}.
\label{eq:mond_frequency_TF_explicit}
\end{equation}
This yields the scaling law $\tilde{\Omega}_M^{\mathrm{TF}} \propto \beta^{-1/3} \eta^{5/6}$. In particular, for a fixed $\eta$, the frequency scales as $\tilde{\Omega}_M \propto \beta^{-1/3}$, a much slower decay with increasing interaction strength compared to the Newtonian case. For the Newtonian potential, the TF approximation fails, and no simple analytical scaling for $\tilde{\Omega}_N$ exists; its dependence on $\beta$ must be determined numerically.

\subsection{Dynamical Evolution and Numerical Results}

\begin{figure}[!ht]
\centering
\includegraphics[width=0.45\textwidth]{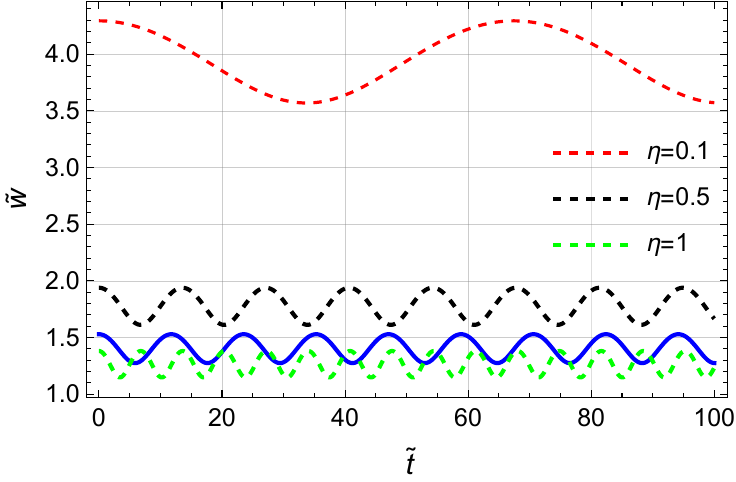}
\caption{Width oscillations for $\beta=1$. Newtonian potential (blue solid). MOND potential for $\eta=0.1$ (red dashed), $\eta=0.5$ (black dashed), and $\eta=1$ (green dashed). The dashed horizontal lines indicate the respective equilibrium widths from Table~\ref{tab:ground_state_properties}.}
\label{fig:width_evolution_comparison}
\end{figure}

The full dynamical behavior is explored by numerically solving Eqs.~(\ref{eq:dimless_newton_width_eq}) and (\ref{eq:dimless_mond_width_eq}) using the equilibrium widths from Table~\ref{tab:ground_state_properties}. For $\beta=1$, the equilibrium widths are $\tilde{w}_{0N} \approx 1.39$, $\tilde{w}_{0M}(\eta=0.1) \approx 3.90$, $\tilde{w}_{0M}(\eta=0.5) \approx 1.76$, and $\tilde{w}_{0M}(\eta=1) \approx 1.26$. Starting from a $10\%$ perturbation, $\tilde{w}(0)=1.1\tilde{w}_0$ with $\dot{\tilde{w}}(0)=0$, we obtain the width oscillations shown in Fig.~\ref{fig:width_evolution_comparison}.

The simulations confirm the theoretical analysis. The oscillation frequency is lowest in the deep-MOND regime ($\eta=0.1$) and increases with $\eta$, with the Newtonian frequency lying between the MOND frequencies for $\eta=0.5$ and $\eta=1$. This trend directly reflects the $\eta$-dependence in Eq.~(\ref{eq:mond_frequency_final}). Furthermore, the oscillations in the MOND potential exhibit noticeable nonlinear distortion, especially at larger $\eta$, due to the non-harmonic logarithmic potential. In contrast, Newtonian oscillations are nearly sinusoidal. The average oscillation widths align well with the equilibrium values from Table~\ref{tab:ground_state_properties}, verifying the self-consistency of the model.

\subsection{Experimental Implications}

The distinct frequency behaviors provide a clear experimental signature to differentiate between the gravitational models. The collective breathing mode can be excited by a sudden change in the trap or interaction strength. Measuring the condensate width over time yields the oscillation frequency $\tilde{\Omega}$ as a function of $\beta$.

For a MOND-type logarithmic potential, a plot of $\ln\tilde{\Omega}$ versus $\ln\beta$ is expected to approach a constant slope of approximately $-1/3$ in the strong interaction regime (from $\tilde{\Omega}_M \propto \beta^{-1/3}$). In contrast, for a Newtonian $1/r$ potential, the plot will not show a constant slope due to the absence of a simple scaling law. For a typical $^{87}\text{Rb}$ BEC experiment with $N \sim 10^5$ and $l_0 \sim 1\ \mu\text{m}$, $\beta$ can be tuned across $10^2\sim10^4$ via Feshbach resonance, providing a sufficient range to test these predictions.
\section{Conclusion}\label{co}
We have investigated the ground-state properties and collective dynamics of a Bose-Einstein condensate under two spherically symmetric gravitational potentials: the Newtonian potential and a logarithmic potential inspired by the deep-MOND regime. Using a Gaussian variational approach for the ground state and a time-dependent variational method for the dynamics, we derived self-consistent equations that describe the condensate's equilibrium width, energy, and monopole oscillation frequency.

Our analysis reveals that the MOND parameter $\eta$ plays a decisive role in determining the condensate's properties. In the ground state, bound solutions exist for the MOND potential only in the deep-MOND regime ($\eta \ll 1$), where the condensate is significantly larger and more weakly bound than its Newtonian counterpart. As the repulsive interaction strength $\beta$ increases, the equilibrium width grows in both potentials, but with fundamentally different scaling behaviors. In the strong-interaction limit, the MOND potential exhibits a clean scaling law $\tilde{\sigma}_0^M \propto \beta^{1/3}$, which is accurately captured by the TF approximation. In contrast, the Newtonian potential does not admit a simple scaling law because the kinetic energy remains comparable to the gravitational energy even for large $\beta$, causing the TF approximation to fail.

For the collective dynamics, we derived explicit expressions for the monopole oscillation frequencies. The frequency of the MOND-bound condensate is lower than that of the Newtonian case, reflecting the weaker logarithmic confinement. This frequency decreases with increasing $\beta$, and in the strong-interaction limit it scales as $\tilde{\Omega}_M \propto \beta^{-1/3}$ for fixed $\eta$. Numerical simulations of the width evolution confirm these trends and further show that oscillations in the MOND potential exhibit noticeable nonlinear distortion due to the non-harmonic form of the logarithmic potential, whereas Newtonian oscillations are nearly sinusoidal.

The contrasting scaling laws in both the equilibrium size ($\beta^{1/3}$ versus no simple scaling) and the oscillation frequency ($\beta^{-1/3}$ versus non-power-law dependence) provide a clear experimental pathway to distinguish between the two gravitational models in quantum simulation experiments. Using Feshbach resonance to tune the s-wave scattering length, the interaction parameter $\beta$ can be varied over several orders of magnitude. Measuring the condensate size and the monopole oscillation frequency as functions of $\beta$ and extracting the scaling exponents offer direct tests of the underlying potential.

Future directions include extending the analysis to more accurate trial wavefunctions, studying the crossover region ($\eta \sim 1$) in detail to map out the critical boundary for bound-state formation, incorporating finite-temperature effects, and designing specific experimental setups (e.g., using optical potentials) to realize the required logarithmic potential in the laboratory.

\paragraph*{Acknowledgments}
This work was supported by the Open Fund of Key Laboratory of Multiscale Spin Physics (Ministry of Education), Beijing Normal University (Grant No.SPIN2024N03), and the Scientific Research Startup Foundation for High-Level Talents at Anqing Normal University (Grant No.241042).

\bibliography{refs}
\bibliographystyle{elsarticle-num-names}

\end{document}